\documentclass[useAMS,usenatbib]{mn2e}

\usepackage{caption}
\usepackage{graphicx}
\usepackage{textcomp}
\usepackage{latexsym}
\usepackage{varioref}
\usepackage{xspace}
\usepackage{makeidx}
\usepackage{verbatim}
\usepackage{tabularx}
\usepackage{epstopdf}
\usepackage{amsmath}
\usepackage{array}
\usepackage{rotating}
\usepackage{hyperref}
\usepackage{float}

\def\Kepler{\textit{Kepler}}

\title[\Kepler\ observations of KIS J1927]
{\Kepler\ observations of the eclipsing cataclysmic variable KIS J192748.53$+$444724.5}
\author[S. Scaringi \textit{et al.}]
{S. Scaringi$^{1}$\thanks{E-mail: simone.scaringi@ster.kuleuven.be}, P.J. Groot$^{2}$, M. Still$^{3,4}$\\ 
$^{1}$Instituut voor Sterrenkunde, K.U. Leuven, Celestijnenlaan 200D, B-3001 Leuven, Belgium \\
$^{2}$Department of Astrophysics/IMAPP, Radboud University Nijmegen, P.O. Box 9010, 6500 GL Nijmegen, The Netherlands \\ 
$^{3}$NASA Ames Research Center, Moffett Field, CA 94035, USA  \\ 
$^{4}$Bay Area Environmental Research Institute, Inc., 560 Third Street West, Sonoma, CA 95476, USA  \\ 
}
\begin{document} 

\date{}

\pagerange{\pageref{firstpage}--\pageref{lastpage}} \pubyear{2013}

\maketitle

\label{firstpage}

\begin{abstract}
We present results from long cadence \Kepler\ observations covering $97.6$ days of the newly discovered eclipsing cataclysmic variable KIS J192748.53$+$444724.5/KIC 8625249. We detect deep eclipses of the accretion disk by the donor star every 3.97 hours. Additionally, the \Kepler\ observations also cover a full outburst for this cataclysmic variable, making  KIS J192748.53$+$444724.5 the second known eclipsing cataclysmic variable system in the \Kepler\ field of view. We show how in quiescence a significant component associated to the hot-spot is visible preceding the eclipse, and that this component is swamped by the brightness increase during the outburst, potentially associated with the accretion disk. Furthermore we present evidence for accretion disk radius changes during the outburst by analysing the out-of-eclipse light levels and eclipse depth through each orbital cycle. We show how these parameters are linearly correlated in quiescence, and discuss how their evolution during the outburst is 
suggesting disk radius changes and/or radial temperature gradient variations in the disk. 

\end{abstract}


\section{Introduction}

Cataclysmic variables (CVs) are interacting close binary systems where a late-type star transfers material to a white dwarf (WD) companion via Roche lobe overflow. With a system orbital period in the range of hours, the transferred material from the secondary star forms an accretion disk surrounding the WD. As angular momentum is transported outwards in the disk, material will approach the inner-most regions close to the WD in the absence of strong magnetic fields, and eventually accrete onto the compact object. Eclipsing CVs are particularly useful not only because the system parameters can be recovered (such as the masses of the two stars), but also because modelling the eclipses allows us to study in great detail the physics of the accretion disk (see \citealt{horne85,feline04}). In total, $208$ eclipsing systems are known (\citealt{RKcat} version 7.12). In this \textit{Letter} we report on the discovery of an eclipsing dwarf-nova type CV within the \Kepler\ field-of-view: KIC 8625249/KIS J192748.53$+$444724.5 (hereafter KIS J1927). This is the second know eclipsing CV in the \Kepler\ field, after V447 Lyr (\citealt{ramsay}).

\begin{figure*}
\includegraphics[width=1\textwidth]{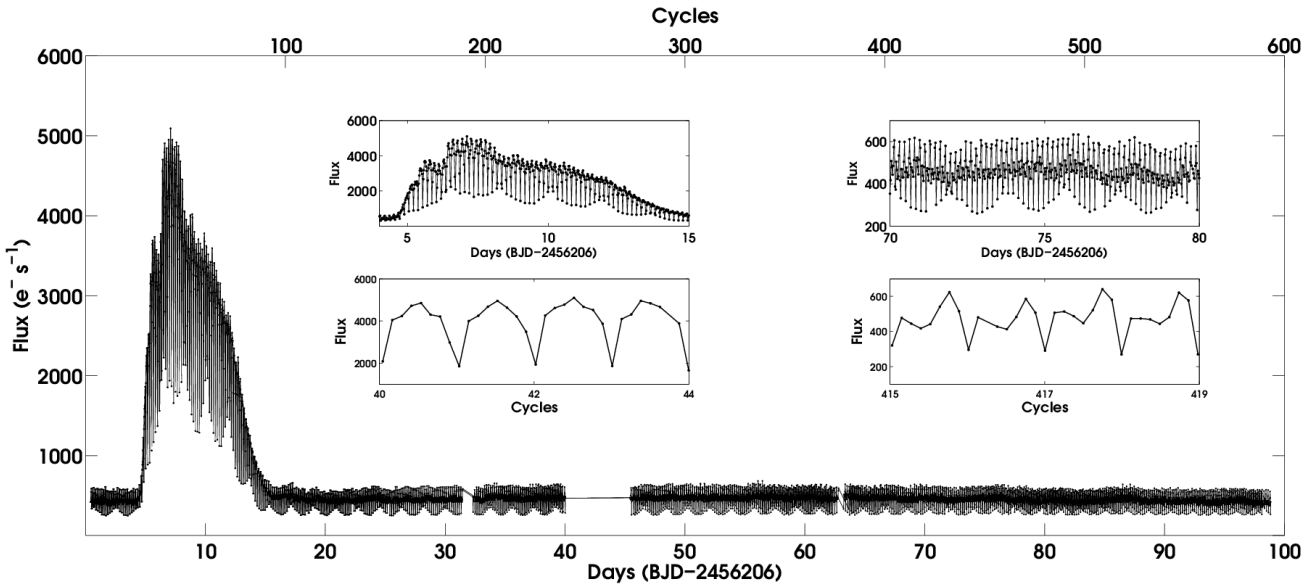}
\caption{Long cadence \Kepler\ Quarter 15 lightcurve of KIS J1927. The top axis displays the orbital cycle. The left-most insets display zoomed in versions of the lightcurve during the outburst, whilst the right-most insets display quiescent parts. The data points have been connected by lines to guide the eye.}
\label{fig:1}
\end{figure*}

KIS J1927 was first discovered as a CV by \cite{scaringi_ID} via spectroscopic follow-up of sources from the \Kepler-INT Survey (\citealt{greiss}) displaying both H$\alpha$ and blue colour excess. Soon after the discovery, the \Kepler\ satellite began monitoring this object (\Kepler\ passband magnitude of $Kp=18.4$) with a timing resolution of $29.4$ minutes (long cadence mode). Only 7 CVs have been observed and studied with \Kepler\ (\citealt{still10,cannizzo10,wood,cannizzo12,barclay,scaringi_rms,scaringi_qpo,ramsay,osaki12,kato13,osakikato13,katoosaki13}), but an additional 8 have recently been monitored. 

In Section \ref{sec:data} we introduce the \Kepler\ satellite and the available observations, and present the orbital lightcurve of KIS J1927. In section \ref{sec:timing} we provide an ephemeris for the system, whilst in section \ref{sec:folded} we discuss the folded lightcurve on the orbital period. Section \ref{sec:disk} discusses how the \Kepler\ photometry provides evidence for accretion disk radius changes and/or radial temperature gradient variations during the observed outburst of KIS J1927, and places our observations in the context of previous work. Our conclusions are drawn in section \ref{sec:conclusion}, and prospects for future \Kepler\ observations of KIS J1927 are discussed.

\section{Observations}\label{sec:data}

The \Kepler\ mission's primary science objective is to discover Earth-sized planets in the habitable zone of Sun-like stars (\citealt{borucki,haas,koch}). The spacecraft is in an Earth-trailing orbit allowing it to continuously monitor the same field-of-view (FOV). The shutterless photometer (with a response function covering the wavelength range $4000-9000$ \AA) has a 116 deg$^2$ FOV and makes use of 6.02 second integrations (plus an additional $0.52$ seconds for CCD readout). Only pixels containing pre-selected targets are saved due to telemetry bandwidth and onboard memory constraints. Up to 170,000 targets can be observed in long cadence (LC) mode, where 270 integrations are summed onboard the spacecraft for an effective 29.4 minute exposure, and up to 512 targets can be observed in short cadence (SC) mode, where nine integrations are summed for an effective 58.8 second exposure. Gaps in the photometric lightcurves are the result of quarterly data downlinks, as well as \Kepler\ occasionally entering anomalous safe modes. During such events no data are recorded and for a few days following these events the data are always correlated due to the spacecraft not being in thermal equilibrium. Further details of artefacts within \Kepler\ light curves can be found in the Kepler Data Release Notes 20 (\citealt{DRN20}). Here, we make no attempt to correct these artefacts, but simply remove them from the light curve.

\begin{figure}
\includegraphics[width=0.45\textwidth]{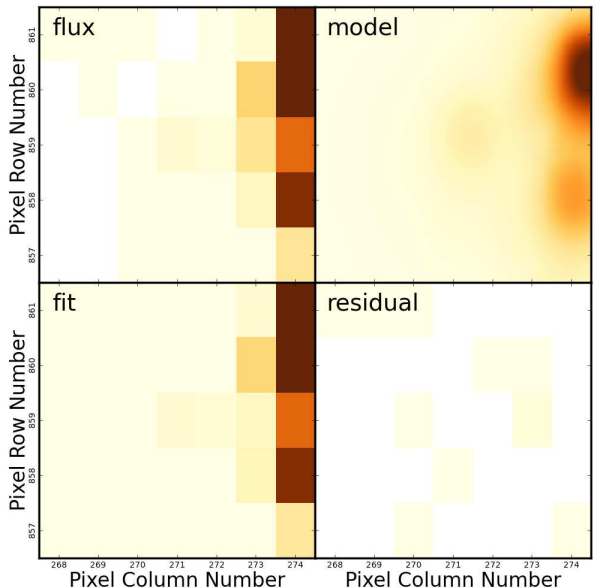}	
\caption{A typical PSF photometry fit to the pixels collected from a single time stamp. A colour version of this figure is available online.}
\label{fig:2}
\end{figure}

The data for KIS J1927 discussed in this \textit{Letter} is that of the first quarter with available and reduced observations (Quarter 15: 04 October 2012 - 06 January 2013) obtained in LC mode. KIS J1927 resides in a crowded field with a number of close neighbours identified, including the $Kp = 17.4$ object KIC 8625243. The archived Simple Aperture Photometry is based upon the summation of two collected pixels with CCD module 13.3 coordinates (271,859) and (272,859). We expect these two pixels to be contaminated by the Point-spread-function (PSF) wings of near neighbours, and the correction for contamination in the archived Pre-search Conditioning Data is overly simplified, resulting in the eclipse depths to be underestimated in the archived Simple Aperture Photometry. To rectify this situation we extract new photometry from the archived target pixels using PSF photometry. The PSF model was downloaded from the Mikulski Archive for Space Telescopes (MAST\footnote{\url{http://archive.stsci/edu/kepler/fpc.html}}). A more precise PSF distribution was obtained by interpolation over the position of KIS J1927 and this model was fit to the target pixels, at each photometric time stamp. A fit adopting three significant sources within the target mask proved sufficient to reduce the residuals to an acceptable level (Pearson's $\chi^2$ = 173 for 26 degrees of freedom). The resulting photometric time series for the target star is provided in Fig. \ref{fig:1}. A typical fit to the pixels collected from a single time stamp is provided in Fig. \ref{fig:2}. The median eclipse depth (relative to the out-of-eclipse light) within the Simple Aperture Photometry was $90.3$ electrons/second, whereas the the median eclipse depth within the PSF photometry is 161.2 electrons/second.

The lightcurve in Fig. \ref{fig:1} shows a large amplitude outburst starting shortly after the beginning of the observations, similar to the outbursts observed in dwarf-nova type CVs (see \citealt{still10,cannizzo10,cannizzo12,wood,kato13}). The Fourier transform of the quiescent interval (BJD$>2456226$ days) is shown in Fig. \ref{fig:2}, where the orbital period is clearly visible, as well as higher harmonics due to the non-sinusoidal lightcurve shape through each orbit. Also alias periods are present, which are due to the close proximity of integer multiples of the sampling frequency to the system orbital period. The most notable aspect of the lightcurve are the eclipses occurring every 3.97 hours as the binary inclination angle is high enough to cause the donor star to eclipse the accreting white dwarf and/or its associated accretion disk and hot spot. 

\begin{figure}
\includegraphics[width=0.45\textwidth]{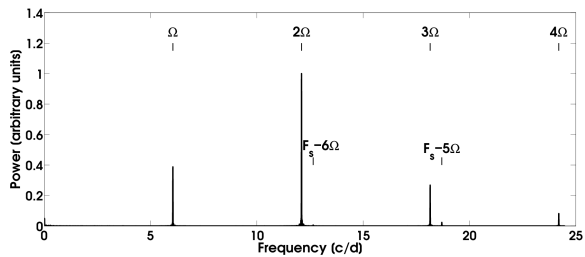}	
\caption{Fourier transform of the long cadence \Kepler\ Quarter 15 lightcurve of KIS J1927 during quiescence. $\Omega$ marks the orbital frequency (with $2\Omega$, $3\Omega$ and $4\Omega$ marking the higher harmonics), whilst $F_s-6\Omega$ and $F_s-5\Omega$ mark two detected aliasing frequencies, where $F_s$ is the sampling frequency.}
\label{fig:3}
\end{figure}

\section{Eclipse Timings}\label{sec:timing}
We estimated the arrival times of the eclipse minima by fitting a spline function independently to each orbital cycle, and determining the time of minimum flux within the fitted curve. In order to obtain the eclipse ephemeris we then fitted a linear curve to the observed mid-eclipse times for every observed cycle. The accuracy of the ephemeris is then assumed to be the small scatter around the fit, which is mainly caused by the aliasing of the sampling frequency to the orbital period. The mid-eclipse ephemeris is then:
\begin{equation}
BJD_{min}= 2456206.0845(17)  +  0.1653077(49)\cdot N,
\label{eq:1}
\end{equation}
where $N$ is the cycle number. The $1\sigma$ uncertainty for the parameters are given in parentheses for the last digits.

\section{Folded lightcurve}\label{sec:folded}

\begin{figure}
\includegraphics[width=0.45\textwidth]{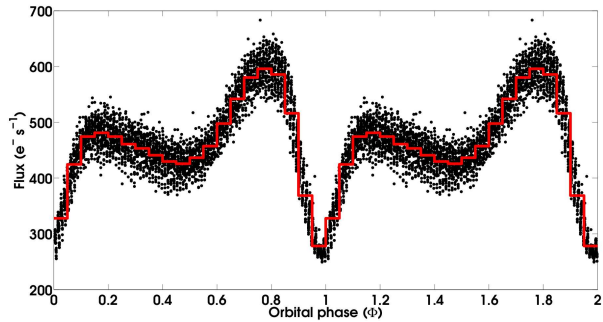}	
\caption{Folded lightcurve of KIS J1927 on 3.97 hours with data obtained with the \Kepler\ satellite in LC mode during quiescence (BJD$>2456226$ days). The solid red line shows the folded lightcurve with 20 phase bins. A coloured version of this figure is available online.}
\label{fig:4}
\end{figure}

Fig. \ref{fig:4} shows the lightcurve folded on the orbital period of 3.97 hours during the quiescent interval (BJD$>2456226$ days). The median eclipse depth relative to the out-of-eclipse flux is $34\%$. However, this value is most-likely underestimated as a result of the data cadence. Future observations with a faster cadence may better resolve each orbital cycle and reveal the eclipses to be deeper than what is observed here. The increase in flux at the orbital phase just preceding the eclipse ($\phi\approx0.8-0.9$, where $\phi$ is the orbital phase) can be explained by the hot-spot (where the accretion stream from the donor star impacts the outer-edges of the accretion disk) being observed nearly face on (\citealt{wood86}). It is also possible that the hot-spot itself is also being eclipsed by the donor star. This would explain the observed flux descent after the eclipse ($\phi\approx0.1-0.3$) as a continuation of the hot-spot emission, and would imply that the maximum hot-spot brightness occurs during the eclipse.

Assuming we can derive the mass ratio of the system, we can place constraints on the system inclination. To do this we use the semi-empirical donor tracks of \cite{knigge_donor} to infer the mass of the secondary star of $M_{2}=0.32M_{\odot}$ from the observed orbital period. Since the mass ratio $q=M_{2}/M_{1}$ has to be smaller than $2/3$ for stable mass transfer to occur (\citealt{warner}), and since the mass of the primary can at most be the Chandrasekhar limit ($1.4M_{\odot}$), we infer a mass ratio range of $0.23<q<0.67$. We note however that the mass ratio might be larger than 0.35 for this system as no superhumps are detected during outburst (although this might be caused by the low sampling rate), but we employ a larger conservative range. By fitting a Gaussian function to the folded eclipse profile (between $0.85<\phi<1.1$) we also infer a full width at half maximum of the eclipse of $\Delta\phi=0.095$. We then employ the method of \cite{horne85} to deduce an inclination of $i>80^{o}$.

\section{Varying accretion disk radius}\label{sec:disk}

Fig. \ref{fig:1} shows the \Kepler\ lightcurve of KIS J1927 as a function of orbital cycle, with insets displaying zoomed portions of the lightcurve during outburst and quiescence. It is clear from Fig. \ref{fig:1} that there are significant changes in the eclipse profiles as the system switches between outburst and quiescence. Most notably, the increase in flux at $\phi\approx0.8-0.9$ associated with the hot-spot is much less pronounced during the outburst. This is potentially suggesting that during the outburst, the optical light is dominated by the accretion disk and the bright spot makes a significantly smaller contribution to the total optical flux as compared to quiescence, and/or that the hot-spot emitting region has changed from a small, compact, region to a larger structure over the disk. Additionally, the eclipse depth relative to the out-of-eclipse light changes from $34\%$ in quiescence to $58\%$ in outburst.

\begin{figure}
\includegraphics[width=0.5\textwidth]{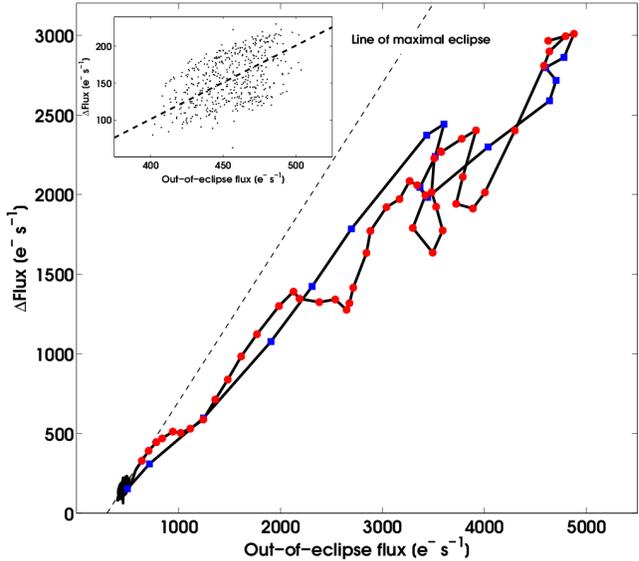}	
\caption{The eclipse depth in KIS J1927 as a function of out-of-eclipse light. Marked in black dots are measurements in quiescence, blue squares are outburst rise and red circles are outburst decay cycles. The thick black line joins consecutive cycles. A zoomed version of the quiescent observations is displayed in the top-left inset, also showing the straight line fit representing the ``Line of maximal eclipse'' (see section \ref{sec:disk}). A coloured version of this figure is available online.}
\label{fig:5}
\end{figure}

The \Kepler\ observations of KIS J1927 displayed an increase in brightness of $2.5$ magnitudes during the outburst. Fig. \ref{fig:5} shows the relation between the out-of-eclipse brightness and the eclipse depth. Studies on how these variations are correlated in eclipsing CVs have been presented by \cite{gs_pav} and \cite{walker}. These studies found that during quiescence the out-of-eclipse light is linearly correlated with the eclipse depth, but also that deviations are observed from this correlation, namely the ``Walker branch'' (as these excursions were first noted by \citealt{walker}) and the ``Shallow branch'' (naming after Fig.~2 of \citealt{gs_pav}). The explanation for the linear correlation between the out-of-eclipse light and the eclipse depth is simple geometry, where the same part of the disk is eclipsed during every orbital cycle, assuming the radial temperature gradient of the disk remains constant. If the accretion disk brightens by say one magnitude (but does not increase in size), then the eclipse depth will also increase by the same amount. In principle, during each cycle, the secondary star may totally eclipse the accretion disk, and the correlation line has been named the ``Line of Totality''. However a total eclipse of the accretion disk is not a necessity to produce a linear correlation between the out-of-eclipse light and the eclipse depth: all that is required is that the same fraction of the disk be eclipsed during each orbital cycle, and we thus rename this line the ``Line of maximal eclipse''. 

We have performed a similar analysis as that presented by \cite{gs_pav}. For each orbital cycle we measured the out-of-eclipse light and eclipse depth as discussed in section \ref{sec:folded}, and show the out-of-eclipse light as a function of eclipse depth for KIS J1927 in Fig. \ref{fig:5}. The ``Line of maximal eclipse'' has been derived from fitting a linear relation to observations in quiescence (BJD$>2456226$ days) by keeping the gradient fixed to unity and only allowing the intercept to vary, as expected from the ``Line of maximal eclipse'' (see top-left inset of Fig. \ref{fig:5}). To test whether the assumption of unity gradient is consistent with the data we performed a two-parameter fit, allowing both the gradient and intercept to vary. We assigned counting (Poisson) errors to the out-of-eclipse flux axis, whilst we summed in quadrature those for the eclipse depth, taking into account the covariance term as the two parameters are not independent. We recover a gradient of $1.07\pm0.14$, consistent with unity. We thus fix the gradient to unity and obtain an out-of-eclipse flux intercept is $299.1\pm2.7$ electrons/second, which, in the case of a total eclipse, would represent the brightness of the secondary star of $Kp=18.9$. The line represents the maximal fractional eclipse depth allowed by the system, which occurs during quiescence when the largest relative disk fraction is eclipsed. The outburst rise is marked with blue squares, whilst the decline with red circles, and the thick line joins consecutive orbital cycles. Several apparent ``branches'' are also present in Fig. \ref{fig:5}, however these seem to be correlated with the aliasing pattern, causing the eclipse depth to be systematically shallower at regular intervals. However, from Fig. \ref{fig:5} it can be seen how the observed tracks immediately deviate from the ``Line of maximal eclipse'' at the outburst onset. 

During the outburst evolution, a decrease in the radial temperature gradient (say from $T\approx R^{0}$ to $T\approx R^{-3/4}$) is expected. This will cause observations to lie above the ``Line of maximal eclipse'' in the case where the inner-most edges of the disk are being eclipsed during each cycle.  The only way to then explain our observations is by increasing the radial size of the disk during the outburst. If the inner-most edges of the disk are not being eclipsed, then an increase in the radial temperature gradient could explain our observations. However we note that CV accretion disks are theoretically expected to expand in radius, and the temperature gradient is theoretically expected to decrease, during outbursts (\citealt{FKR, anderson}). Furthermore, many CVs display this expansion in their observational properties (e.g. U Gem: \citealt{smak} and Z Cha: \citealt{ODon}).

\section{Conclusion}\label{sec:conclusion}

We have reported on long cadence \Kepler\ observations of the eclipsing CV KIS J1927. The system has an orbital period of 3.97 hours, and displayed a $\approx2.5$ magnitude, $\approx10$ day-long, outburst during the 97.6 day-long observation. The quiescent folded lightcurve displays a significant contribution from the hot spot, which is swamped during the outburst, potentially by the bright accretion disk. We have also reported on evidence for accretion disk radius changes during the outburst of this system by studying the out-of-eclipse light levels versus the eclipse depth during the outburst. We find that the eclipse depth and out-of-eclipse light levels are linearly correlated in quiescence, and that this correlation is offset during outburst. This result is suggesting that the accretion disk is increasing in radius and/or temperature during the outburst. Similar implications on the other eclipsing CV in the \Kepler\ field (V447 Lyr, \citealt{ramsay}) have also been deduced, as well as previous work using ground based observations (\citealt{gs_pav,rutten}).

The main drawback from our analysis has been the lightcurve cadence, which coarsely samples the eclipses. Short-cadence \Kepler\ observations of this object, with a sampling timescale of 58.5 seconds, would potentially allow to model the lightcurve via eclipse mapping, and to separate out each individual contribution of the lightcurve for every orbital cycle (accretion disk, hot spot, white dwarf and secondary star). Furthermore, eclipse mapping on short cadence data will allow us to locate the physical origin of flickering in CVs (\citealt{baptista04,scaringi_rms}) with unprecedented precision. Additionally, eclipse mapping of this system using short cadence data will potentially allow us to track the disk evolution during the outburst, and to infer both radius as well as temperature changes during the rise and fall of the outburst. If additional observations by \Kepler\ become viable after investigation of further 2- or 3- reaction wheel operations, KIS J1927 is a valuable potential target for short cadence observations.

\section*{Acknowledgements}
This research has made use of NASA's Astrophysics Data System Bibliographic Services. S.S. acknowledges funding from the FWO Pegasus Marie Curie Fellowship program. Additionally, S.S. acknowledges the use of the astronomy \& astrophysics package for Matlab (Ofek in prep.).

\bibliographystyle{mn}
\bibliography{KIC86_paper}

\begin{thebibliography}{32}
\expandafter\ifx\csname natexlab\endcsname\relax\def\natexlab#1{#1}\fi

\bibitem[{{Anderson}(1988)}]{anderson}
{Anderson} N., 1988, \apj, 325, 266

\bibitem[{{Baptista} \& {Bortoletto}(2004)}]{baptista04}
{Baptista} R., {Bortoletto} A., 2004, \aj, 128, 411

\bibitem[{{Barclay} {et~al.}(2012){Barclay}, {Still}, {Jenkins}, {Howell}, \&
  {Roettenbacher}}]{barclay}
{Barclay} T., {Still} M., {Jenkins} J.~M., {Howell} S.~B., {Roettenbacher}
  R.~M., 2012, \mnras, 422, 1219

\bibitem[{{Borucki} {et~al.}(2010){Borucki}, {Koch}, {Basri}, {Batalha},
  {Brown}, {Caldwell}, {et~al.}}]{borucki}
{Borucki} W.~J., {Koch} D., {Basri} G., {Batalha} N., {Brown} T., {Caldwell}
  D., {et~al.}, 2010, Science, 327, 977

\bibitem[{{Cannizzo} {et~al.}(2012){Cannizzo}, {Smale}, {Wood}, {Still}, \&
  {Howell}}]{cannizzo12}
{Cannizzo} J.~K., {Smale} A.~P., {Wood} M.~A., {Still} M.~D., {Howell} S.~B.,
  2012, \apj, 747, 117

\bibitem[{{Cannizzo} {et~al.}(2010){Cannizzo}, {Still}, {Howell},
  {et~al.}}]{cannizzo10}
{Cannizzo} J.~K., {Still} M.~D., {Howell} S.~B., {et~al.}, 2010, \apj, 725,
  1393

\bibitem[{{Feline} {et~al.}(2004){Feline}, {Dhillon}, {Marsh}, \&
  {Brinkworth}}]{feline04}
{Feline} W.~J., {Dhillon} V.~S., {Marsh} T.~R., {Brinkworth} C.~S., 2004,
  \mnras, 355, 1

\bibitem[{{Frank} {et~al.}(2002){Frank}, {King}, \& {Raine}}]{FKR}
{Frank} J., {King} A., {Raine} D.~J., 2002, {Accretion Power in Astrophysics:
  Third Edition}. Cambridge Press

\bibitem[{{Greiss} {et~al.}(2012){Greiss}, {Steeghs}, {G{\"a}nsicke},
  {Mart{\'{\i}}n}, {Groot}, {Irwin}, {et~al.}}]{greiss}
{Greiss} S., {Steeghs} D., {G{\"a}nsicke} B.~T., {Mart{\'{\i}}n} E.~L., {Groot}
  P.~J., {Irwin} M.~J., {et~al.}, 2012, \aj, 144, 24

\bibitem[{{Groot} {et~al.}(1998){Groot}, {Augusteijn}, {Barziv}, \& {van
  Paradijs}}]{gs_pav}
{Groot} P.~J., {Augusteijn} T., {Barziv} O., {van Paradijs} J., 1998, \aap,
  340, L31

\bibitem[{{Haas} {et~al.}(2010){Haas}, {Batalha}, {Bryson}, {Caldwell},
  {Dotson}, {Hall}, {et~al.}}]{haas}
{Haas} M.~R., {Batalha} N.~M., {Bryson} S.~T., {Caldwell} D.~A., {Dotson}
  J.~L., {Hall} J., {et~al.}, 2010, \apjl, 713, L115

\bibitem[{{Horne}(1985)}]{horne85}
{Horne} K., 1985, \mnras, 213, 129

\bibitem[{{Kato} \& {Maehara}(2013)}]{kato13}
{Kato} T., {Maehara} H., 2013, 1303.1237

\bibitem[{{Kato} \& {Osaki}(2013)}]{katoosaki13}
{Kato} T., {Osaki} Y., 2013, 1305.5636

\bibitem[{{Knigge} {et~al.}(2011){Knigge}, {Baraffe}, \&
  {Patterson}}]{knigge_donor}
{Knigge} C., {Baraffe} I., {Patterson} J., 2011, \apjs, 194, 28

\bibitem[{{Koch} {et~al.}(2010){Koch}, {Borucki}, {Basri}, {et~al.}}]{koch}
{Koch} D.~G., {Borucki} W.~J., {Basri} G., {et~al.}, 2010, \apjl, 713, L79

\bibitem[{{O'Donoghue}(1986)}]{ODon}
{O'Donoghue} D., 1986, \mnras, 220, 23P

\bibitem[{{Osaki} \& {Kato}(2012)}]{osaki12}
{Osaki} Y., {Kato} T., 2012, 1212.1516

\bibitem[{{Osaki} \& {Kato}(2013)}]{osakikato13}
---, 2013, 1305.5877

\bibitem[{{Ramsay} {et~al.}(2012){Ramsay}, {Cannizzo}, {Howell}, {Wood},
  {Still}, {Barclay}, \& {Smale}}]{ramsay}
{Ramsay} G., {Cannizzo} J.~K., {Howell} S.~B., {Wood} M.~A., {Still} M.,
  {Barclay} T., {Smale} A., 2012, \mnras, 425, 1479

\bibitem[{{Ritter} \& {Kolb}(2003)}]{RKcat}
{Ritter} H., {Kolb} U., 2003, \aap, 404, 301

\bibitem[{{Rutten} {et~al.}(1992){Rutten}, {Kuulkers}, {Vogt}, \& {van
  Paradijs}}]{rutten}
{Rutten} R.~G.~M., {Kuulkers} E., {Vogt} N., {van Paradijs} J., 1992, \aap,
  265, 159

\bibitem[{{Scaringi} {et~al.}(2013){Scaringi}, {Groot}, {Verbeek}, {Greiss},
  {Knigge}, \& {K{\"o}rding}}]{scaringi_ID}
{Scaringi} S., {Groot} P.~J., {Verbeek} K., {Greiss} S., {Knigge} C.,
  {K{\"o}rding} E., 2013, \mnras, 428, 2207

\bibitem[{{Scaringi} {et~al.}(2012{\natexlab{a}}){Scaringi}, {K{\"o}rding},
  {Uttley}, {Groot}, {Knigge}, {Still}, {et~al.}}]{scaringi_qpo}
{Scaringi} S., {K{\"o}rding} E., {Uttley} P., {Groot} P.~J., {Knigge} C.,
  {Still} M., {et~al.}, 2012{\natexlab{a}}, \mnras, 427, 3396

\bibitem[{{Scaringi} {et~al.}(2012{\natexlab{b}}){Scaringi}, {K{\"o}rding},
  {Uttley}, {Knigge}, {Groot}, \& {Still}}]{scaringi_rms}
{Scaringi} S., {K{\"o}rding} E., {Uttley} P., {Knigge} C., {Groot} P.~J.,
  {Still} M., 2012{\natexlab{b}}, \mnras, 421, 2854

\bibitem[{{Smak}(1984)}]{smak}
{Smak} J., 1984, \actaa, 34, 93

\bibitem[{{Still} {et~al.}(2010){Still}, {Howell}, {Wood}, {et~al.}}]{still10}
{Still} M., {Howell} S.~B., {Wood} M.~A., {et~al.}, 2010, \apjl, 717, L113

\bibitem[{{Thompson} {et~al.}(2013){Thompson}, {Christiansen}, {Jenkins},
  {Caldwell}~and, {Bryson}, {Burke}, {Campbell}, {et~al.}}]{DRN20}
{Thompson} S., {Christiansen} J., {Jenkins} J., {Caldwell}~and {Barclay} T.,
  {Bryson} S., {Burke} C., {Campbell} J., {et~al.}, 2013

\bibitem[{{Walker}(1963)}]{walker}
{Walker} M.~F., 1963, \apj, 137, 485

\bibitem[{{Warner}(2003)}]{warner}
{Warner} B., 2003, {Cataclysmic Variable Stars}. Cataclysmic Variable Stars, by
  Brian Warner, pp.~592.~ISBN 052154209X.~Cambridge, UK: Cambridge University
  Press, September 2003.

\bibitem[{{Wood} {et~al.}(1986){Wood}, {Horne}, {Berriman}, {Wade},
  {O'Donoghue}, \& {Warner}}]{wood86}
{Wood} J., {Horne} K., {Berriman} G., {Wade} R., {O'Donoghue} D., {Warner} B.,
  1986, \mnras, 219, 629

\bibitem[{{Wood} {et~al.}(2011){Wood}, {Still}, {Howell}, {Cannizzo}, \&
  {Smale}}]{wood}
{Wood} M.~A., {Still} M.~D., {Howell} S.~B., {Cannizzo} J.~K., {Smale} A.~P.,
  2011, \apj, 741, 105

\end{thebibliography}

\label{lastpage}

\end{document}